# The effect of chirped intense femtosecond laser pulses on the Argon cluster

H. Ghaforyan[1], R. Sadighi[2*], E. Irani[2]

[1] Department of physics, Payame Noor Universtiy, PO BOX 19395-3697 Tehran,Iran

[2] Department of Physics, Sharif University of Technology, P.O. Box 11365-9567, Tehran, Iran

## Abstract:

The interaction of intense femtosecond laser pulses with atomic Argon clusters has been investigated by using nano-plasma model. Based on the dynamic simulations, ionization process, heating and expansion of a cluster after irradiation by femtosecond laser pulses at intensities up to $2\times10^{17}$ $Wcm^{-2}$ are studied. The analytical calculation provides ionization ratefor different mechanisms and time evolution of the density of electrons for different pulse shapes. In this approach the strong dependence of laser intensity, pulse duration and laser shape on the electron energy, the electron density and the cluster size are presented using the intense chirped laser pulses. Based on the presented theoretical modifications, the effect of chirped laser pulse on the complex dynamical process of the interaction is studied. It is found that the energy of electrons and the radius of cluster for the negatively chirped pulsesare improved up to 20% in comparison to the unchirped and positively chirped pulses.

The authors declare that there is no conflict of interest regarding the publication of this paper.

**I. INTRODUCTION:**

[*] Corresponding author: Tel:+982166164526; Fax: +982166022711.
Email address: sadighi@sharif.ir

Developments of ultra-short intense laser pulses through CPA and OPCPA techniqueshave attracted increasing attention in the main subject of laser-matter interaction [1]. These pulses are used for the generation of quasi mono-energetic electron beams up to mega-electron volts [2-6]. The ponderomotive force of laser accelerates plasma electrons to the relativistic energies over several MeV. In recent years, considerable progress has been made in increasing energy and quality of electron beams [7-8].

High-energy electrons have various applications, including fast ignition of fusion reaction [9], production of intense radiation sources such as X-ray, dynamics of photochemistry, ionization and dissociation process, biological and medical technology [10-14]. Optimization and control of this new source of energetic particles with different applications is a subject of current significance which can be modified by unique attractive techniques. One of the efficient attractive methods, interaction of intense laser pulses with large atomic clusters, has opened up several areas of laser plasma science such as electron and ion accelerators [15-17], table-top neutron sources [18], plasma waveguides [19] and X-ray sources [20-22]. Simulation of this interaction is not only scientifically very interesting but also has a wide range of applications. One of the most important applications of laser-cluster interaction is particle acceleration. Low price and low space at the laser-plasma accelerators compared with other accelerators have led to more attention to this type of accelerator.

Although a target cluster which is made by high-pressure gas nozzle shows important properties such as solid-like electron density in some places, the average density of clusters is low. The transmission distance in the cluster is longer than that of in the gas target. In addition, the absorbed laser energy in the cluster is larger than the solid and gas targets [23-24]. Therefore, clusters have unique properties because they have the advantage of both the gas and solid phases.

Another outstanding advantage of clusters in particle accelerations is their capability for controlling the dynamic of interaction of cluster by a laser pulse.

Simulation of laser interaction with large clusters possesses a great challenge. A fully ab initio treatment or molecular dynamics method is not feasible. A good approximation for clusters with an atomic number greater than 10000 atoms is one which considers the cluster as a nano-plasma medium. This model was developed in 1996 by Ditmire and his colleagues [25]. Nano-plasma model successfully justifies all phenomena in the interaction. Several improvements in the nano-plasma models have been made which are in good agreement with experimental results [26]. The pulse shaping techniques and application of chirped pulses have impressively improved controlled laser-plasma coupling [27]. Recently, chirping of short laser pulses has been recognized as a main parameter that can be used to control the dynamics of a system interacting by ultrafast laser pulses. However, there has been much less investigation on the chirp-dependent behavior of dynamics of molecular ionization. Levis and co-workers have recently used optimized laser pulse with intensity of $10^{13}$Wcm$^{-2}$ to control the dissociation patterns in large molecules through various Stark-shifting electronically excited states into resonance condition [28]. Fukuda and co-workers demonstrated that energetic particle emissions from laser-cluster interactions are optimized by manipulating the sign of chirp [29]. Several theoretical models have been proposed to explain the mechanism of the production of highly energetic particles. However, due to the complex dynamical processes of the laser-cluster interaction, a comprehensive model has never been presented to explore the optimal conditions. To the best of our knowledge, there is no theoretical executive report on chirp-dependent behavior of complex dynamics of laser-cluster interaction using nano-plasma model. In this work, a useful theoretical model is used which can properly elucidate the reported experimental results. In order to clarify the laser-cluster interaction, electron and ion density, electron energy and radius of cluster are modified by manipulating the characters of laser fields, such as intensity, pulse duration and chirp parameter via using nano-plasma model.

The aim is to improve the acceleration of particles by considering the modified negatively chirped pulse shapes. Indeed, the effect of chirped pulses creates some modifications on the current calculation models. The effect of chirped pulses in different intensities is also compared. Another feature of this work is to study the time evolution of the density of electrons and ions at different pulse shapes and the results are compared with near transform limited (unchirped) pulses. In addition, the effect of laser intensity, pulse duration and the laser pulse shape have been investigated on the electron energy, charge state of ions and cluster size. When we use the negatively chirped laser pulse, it is found that the energy of electrons and radius of cluster is improved about 20 % and 18% compared to the Gaussian pulse and positively chirped pulse respectively. This is described in detail in the following.

This paper is organized as follows: in Sec. II, the theory of laser-cluster interaction, the mechanisms of cluster ionization, cluster heating, cluster expansion and the ponderomotive force effect are described to provide the effective parameters at laser-cluster interaction; in Sec.III, the simulation results are explained in detail; and the paper is concluded in Sec. IV.

## II. **Theory of laser-cluster interaction**

Various models are used to describe the interaction of lasers with atomic clusters. These models include the coulomb explosion model [30], the ionization ignition model [31], the inner shell excitation [32] and nano-plasma model [25-33]. The most successful method for studying the dynamical evolution of large clusters under strong fields is the nano-plasma model.

In this model, cluster atoms are ionized by the incident laser pulse (inner ionization) and form a nano-plasma sphere. Quasi-free electrons participate in an oscillation that is created by the laser field and interact with other particles. Electron-impact ionization of the cluster produces additional free electrons and vacancies in inner shell which are the origin of the X-ray radiation. Fraction of the electrons finds enough energy to escape from the cluster (external ionization) and leave behind a net positive charge. The cluster, then, expands in response to coulomb explosion and hydrodynamic forces. The enhancement of the laser intensity makes ponderomotive force as an important component of the interaction dynamics. However, there are three main processes (ionization, heating, expansion) in the nano-plasma model that is discussed in detail.

### A. Cluster ionization mechanism

When the rising edge of the laser pulse reaches to the cluster, ionization of the cluster begins and small number of quasi-free electronsis generated. This is called the tunnel ionization and the rate of tunnel ionization is described in this model by ADK formula that is given by [34]

$$W_{ADK}=\mu_a \frac{(2l+1)(l+|m|)!}{2^{|m|}|m|!(l-|m|)!}\left(\frac{2e}{n^*}\right)^{n^*}\frac{1}{2\text{®}n^*}I_p\left(\frac{2E}{\text{®}(2I_p)^{3/2}}\right)^{1/2}\left(\frac{2(2I_p)^{3/2}}{E}\right)^{2n^*-|m|-1}\exp\left[-\frac{2(2I_p)^{3/2}}{3E}\right] \quad (1)$$

Where, the constant e is Euler’s number, $n^*=Z[2I_p]^{-1/2}$ is the effective principal quantum number, $\mu_a$ is the atomic frequency ($\mu_a$= 4.13×10$^{16}$ s$^{-1}$), Ip is the ionization potential of charge state and E is the field of laser in atomic units.

The second ionization mechanism in the cluster occurs in inelastic collisions between electrons and ions. A few electrons produced by optical ionization collide with atoms inside the cluster and create another ionization that is called collisional ionization. The production of higher charge states is dominated by collisional ionization as a result of the high density in the cluster. Rate of collisional ionization in nano-plasma model is calculated by the Lotz equation [35].

$$W_{col} = n_e \frac{a_i q_i}{I_p (kT_e)^{1/2}} \int_{I_p/kT}^{\infty} \frac{e^{-x}}{x} dx \qquad (2)$$

Here, $n_e$ is the electron density, $I_p$ the ionization potential in eV, $q_i$ the number of electron in the outer shell of the ion, and $a_i$ an empirical constant equal to $4.5\times10^{-14}$ $eV^2cm^{-3}$.

In addition to the thermal energy,the electrons in the cluster have a velocity associated with their oscillations in the laser field. Oscillation of the electrons by the laser field leads to another ionization which is given as follows [36]

$$W_{las} \approx n_e \frac{a_i q_i}{2\pi I_p m_e^{1/2} U_p^{1/2}} \left\{ \left[ 3 + \frac{I_p}{U_p} + \frac{3}{32}\left(\frac{I_p}{U_p}\right)^2 \right] \times \ln\left[ \frac{1+\sqrt{1-I_p/2U_p}}{1-\sqrt{1-I_p/2U_p}} \right] - \left( \frac{7}{2} + \frac{3I_p}{8U_p} \right) \times \sqrt{1-I_p/2U_p} \right\} \qquad (3)$$

Hence, the ionization rate of the laser-cluster interaction is equal to

$$W = W_{ADK} + W_{col} + W_{las} \qquad (4)$$

It becomes clear from the results of this research and previous studies that dominant ionization in laser-cluster interaction is collisional ionization. At the same time, three-body recombination occurs and reduces the number of electrons and increases the temperature of the system. The rate of three-body recombination, $\alpha_3$, is [37]

$$\alpha_3 = \frac{4\pi\sqrt{2\pi}}{9} \frac{e^{10} Z^3}{m_e^{1/2} (kT_e)^{9/2}} \ln\sqrt{1+Z^2} \qquad (5)$$

Here, Z is charge state of cluster.

**B. Cluster heating mechanism**

In the hydrodynamic model, the laser deposits energy into the cluster through inverse bremsstrahlung. The plasma is treated as a dielectric medium, and the Drude model gives its dielectric constant as follows:

$$\varepsilon = 1 - \frac{\omega_p^2}{\omega(\omega + i\nu)} \quad (6)$$

The applied electric field is assumed to be uniform all over the plasma (the cluster radius is on the order of 15 nm, and the wavelength of light is 825 nm), and the electric field inside the cluster is given by [38]

$$E = \frac{3}{|\varepsilon + 2|} E_{ext} \quad (7)$$

$E_{ext}$ is laser field outside the cluster. The rate of energy deposition by an applied electric field into a dielectric is given by

$$\frac{\partial U}{\partial t} = \frac{1}{4\pi} \vec{E} \cdot \frac{\partial \vec{D}}{\partial t} \quad (8)$$

By placing electric field inside the cluster, we get

$$\frac{\partial U}{\partial t} = \frac{9\omega}{4\pi} \frac{\mathrm{Im}[E]}{|\varepsilon + 2|^2} |E_{ext}|^2 = 7312.5\omega\, \mathrm{Im}[E] |E|^2 \, eV / \left(nm^3\, fs\right) \quad (9)$$

Where, E is field inside the cluster in atomic unit. The inverse bremsstrahlung (IBS) absorption (or collisional) rate for a clustered plasma is given by

$$Q = \frac{\mu_p^2 \, \mathrm{Re}\, ³(\mu)}{[\mu - \frac{\mu_p^2}{3\mu} - \mathrm{Im}\, ³(\mu)]^2 + [\mathrm{Re}\, ³(\mu)]^2} \frac{|E_0|^2}{8®} \quad (10)$$

Here, $|E_0|^2$ is proportional to the intensity of the laser and $³(\mu)$ the collision frequency.

The electron temperature equation inside the cluster is [39]

$$\frac{\partial T_e}{\partial t} = \frac{2}{3}\frac{Q}{n_e} - \frac{T_e}{n_e}\frac{dn_e}{dt} + \frac{2}{3}\sum_{Z=0}^{\infty} £_i(Z)[\sim_3(Z+1)n_e n_i(Z+1) - S(Z)n_i(Z)] \quad (11)$$

Here, $n_e$ is the electron density and S the collisional ionization rate coefficient that are effective parameters inelectron temperature. After the initial ionization, the electron density in the cluster plasma is much larger than the critical density for 825 nm laser wavelength. However, as the cluster expands, the density falls and the real part of the dielectric constant, which initially has a large negative value, approaches -2.

### C. Cluster expansion mechanism

Cluster expansion is the result of the coulomb pressure and hydrodynamic pressure in the nano-plasma model. Coulomb pressure is the result of the repulsive ions that occur in clusters andcan be found by the following equation [40]

$$P_{coul} = \frac{Q^2 e^2}{8® r^4} \quad (12)$$

Here, **r** is the radius of the cluster. The hydrodynamic pressure created as a result of the expansion of the hot electrons is given by

$P_e = n_e k T_e$ (13)

Here, $n_e$ is the electron density, $T_e$ the electron temperature and $k$ the Boltzmann constant.

### D. Ponderomotive force

Due to effect of the chirped pulses, some modifications are made on the current calculation models. When a beam of high-power laser radiation is used to heat the plasma, radiation pressure becomes significant. Electrons in plasma are accelerated to relativistic energies with ponderomotive force of laser light. In hydrodynamic model, plasma dynamics are driven predominantly by the hydrodynamic pressure and the ponderomotive pressure. The enhancement of the laser intensity at the critical density surface makes ponderomotive forces an important component of the plasma dynamics. The equation of motion of electrons in the presence of an EM wave is described by

$$\vec{F}_L = m\frac{d\vec{v}}{dt} = -e(\vec{E}(\vec{r},t) + \frac{1}{c}\vec{U} \times \vec{B}(\vec{r},t)) \quad (14)$$

$\vec{E}(\vec{r},t)$, $\vec{U}$ and $\vec{B}(\vec{r},t)$ are the electric field, electron velocity and magnetic field, respectively. The electric field is given by

$$\vec{E}(\vec{r},t) = \vec{E}_s(\vec{r}) f(t) \cos \mu_0 t \quad (15)$$

Here, f(t) and $\omega_0$ denote the temporal profile and the carrier frequency, respectively. A Gaussian envelope of *f*(t) is given by

$$f(t) = \exp[-\frac{2\ln 2t^2}{2^2}] \quad (16)$$

Where, τ is FWHM time duration. Linearly chirped laser field can be defined as follows:

$$E(t) = E_0 \exp\{-(1+ib)(\frac{t\sqrt{2\ln 2}}{\tau_L})^2\} \quad (17)$$

b and $\tau_L$ are the chirp parameter and pulse duration. b>0 means a positive chirp and b<0 means a negative chirp. Laser intensity of the chirped laser pulse is expressed as follows:

$$I(t) = I_0 \exp\{[-\frac{t^2}{2\tau^2}][1+\frac{bt}{(t^2+\tau^2)^{1/2}}]^{-1}\} \quad (18)$$

$I_0$ denotes the peak laser intensity of the near transform limited pulse. By using a second-order approximation and expansion of **E** around the $\mathbf{r}_0$, the force exerted on an electron can be obtained as follows:

$$\vec{F}_{NL} = -\frac{\omega_p^2}{8\pi\omega_0^2}\vec{\nabla}\langle\vec{E}^2\rangle \quad (19)$$

Although this force essentially enters the electrons, in the end, it is transmitted to the ions too. When electrons are classified by $\vec{F}_{NL}$, a field separator ($\vec{E}_{es}$) is produced. Therefore, the total force exerted on the electrons would be as follows:

$$\vec{F}_e = -e\vec{E}_{es} + \vec{F}_{NL} \quad (20)$$

Indeed, the ponderomotive force is very important for accelerating the electrons.

## III. Simulation results

Calculations are carried out for Argon clusters under ultra-short laser pulses and three different temporal profile distributions of Gaussian and skewed Gaussian as near transform limited pulse, positively and negatively chirped laser pulse over range of intensities $10^{14}$-$10^{17}$Wcm$^{-2}$. Argon cluster radius is considered 150Å.

The optimal modifications are achieved by adjusting the optimal skew parameter for positive chirp in low intensity about $10^{14}$Wcm$^{-2}$ and negative chirp in high intensity laser pulses more than $2\times10^{15}$Wcm$^{-2}$. The complex dynamics of laser-cluster interaction is modeled by three levels of the ionization, heating, and expansion of a cluster after irradiation by an intense laser pulse. Calculations successfully demonstrate the dependence of the particle density, the electron energy and the radius of cluster with laser parameters and they can be improved by chirped laser pulses. The origin of the time is selected in a way in which $\mathbf{E}_{ext}$ maximum is located at the t=0 with FWHM of 40fs.

Time evolution of the field inside the cluster and the external chirped intense pulse field according to Drude model are shown in Figure 1. This figure denotes that by comparing the electric field inside the cluster and external laser pulse field, an enhanced internal field is achieved which leads to an enhanced ionization rate when the electron density sweeps through $3n_{crit}$. Figure 2 is presented in two dimensions in order to provide a better insight into the importance of the pulse duration and laser intensity effects on the produced electron energy and ion charge states.

Figure 2 elucidates the importance of the pulse duration and laser intensity by studying the manner of charge state and energy of emitted electrons for unchirped laser pulse. It shows that increasing laser intensity and laser pulse duration causes considerable increase of the electrons energy and ion charge states. This can have good impact on many laser–cluster interaction applications due to increasing the electron temperature. The aim is to study the effect of laser parameters including different shapes of laser pulses on the three levels of nano-plasma model. For a proper analysis of time-dependent cluster dynamics, the effective role of electron density in the first stage of this model is justified. According to nano-plasma approach, ionization processes play an important role to strip the electrons and ions to higher charge states. Time evolution of the electron density by 40 fs long near transform limited laser pulses, positively chirped pulses and negatively chirped pulse with peak laser intensity of $10^{17}$ $Wcm^{-2}$ is presented in Figure 3.

.

Based on Figure 3, it is concluded that the ionization starts earlier for negatively chirped pulsein comparison to the unchirped and positively chirped pulses. However, the maximum density of electrons with positively chirped laser pulse is achieved. In addition to different mechanisms of tunneling ionization, collisional ionization and field ionization, recombination of the electrons play an important role in the electron density. Hence, the electron density will be low for negatively chirped pulseas a result of an early start tunneling ionization and the impact of other factors, which is demonstrated in Figure 3. Reducing the electron density with negatively chirped pulse indicates that a strong heating is dominant.

In order to obtain an estimate of the heating and enhancement of the acceleration of particles, the effect of laser intensity and the laser pulse shape as the main effective parameters have been investigated on the electron energy. Time evolution of electron energy for different laser pulse shapes with $1.5\times10^{17}$ $Wcm^{-2}$ laser intensity and the changes of electrons energy at different laser intensities are shown in Figure 4.

When the cluster expands to the resonance condition with time, the laser pulse has already passed. Therefore, for longer duration pulses, the resonant absorption is near the peak of the pulse, which greatly enhances the cluster heating. Figure 4 shows that the enhancement occurs most intensively for negatively chirped pulses and the energy of electrons is improved up to 20 % compared to the Gaussian pulse. However, the results of Figure 3(b) clearly demonstrate that the maximum energy of electrons can be improved by manipulating the character of laser intensity.The electron energy depends on the shape of the laser pulse at different intensities and is increased in condition of negatively chirped pulses when the applied laser pulse is more than $2\times10^{15}Wcm^{-2}$. At low laser intensities, the positive chirp is more effective due to the effect of intensity on the absorption coefficient according to Eq. (10). It should be realized that charging of cluster also causes a radial expansion of the cluster.

Variations of the cluster size by changing the laser intensity at the different laser pulse shapes are shown in Figure 5.

The effective role of the laser intensity in the interaction with clusters is interpreted for cluster to expansion in resonance condition. The dependence of Bremsstrahlung absorption on the laser intensity affects electron density, electron temperature and recombination, resulting in 18% further expansion of the cluster for the negative chirp in comparison to the positive chirp. When the negatively chirped laser pulse interacts with a cluster, ionization starts earlier and heating and expansion process occur quickly and the radius of the cluster will be greater than other states. The density of cluster ions due to cluster expansion and acceleration of ions is declined and followed by enhancing the radius of the cluster. The charge state, ion density and expanding cluster radius are challenging. To obtain the time-dependent cluster radius manner, time evolution of the ion density with the size of cluster at different laser intensities is shown in Figure 6.

Form Figure 6, it is concluded that due to cluster expansion and the ejection of the ions from the cluster, the ions density of cluster decreases with increases in the laser intensities. By increasing the laser intensity, energy absorption in cluster is enhanced and leads to the generation of hot electrons. High energy electrons and ions are ejected and this can reduce the density of cluster ions. As the ion density decreases, the cluster size increases, which is seen from the time needed to reach the asymptotic value. It is more than 5% increase for about 25% decrease of the ion density. Another interesting point of this research work is investigating the role of chirped laser pulses in laser-cluster interaction in various pulse durations.

Figure 7 shows that in short pulse duration, high-energy electrons are mainly generated as a result of cluster interaction with positively chirped laser pulses and for the long laser pulse durations, the role of negatively chirped laser pulses are more effective. This process is due to the effective parameters on the electron temperature as is clarified in Eq.(10).

## IV. Conclusion

In the present work, based on a modified theoretical executive model, behavior of complex dynamics of laser- Argon cluster interaction and its dependence on the laser pulse chirp is analyzed. This useful proposed theoretical model can properly explain the reported experimental results. This proposed theoretical model can properly explain the reported experimental results. In order to clarify the laser-cluster interaction, electron and ion density, electron energy and radius of cluster are modified by manipulating the characters of laser fields, such as intensity, pulse duration and chirp parameter by using nano-plasma model. It should be noted that besides the dependence of the interaction process into the mentioned parameters, there is about 20% improvement for the energy of electrons which are evaluated by negatively chirped femtosecond laser pulse. Based on this model, the dependence of Bremsstrahlung absorption on laser intensity in electron density, electron temperature and recombination indicates that 20% improvement is made for energy of electrons and 18% further expansion of cluster for the negative chirp in comparison to the positive chirp.

**Figure Captions:**

Figure 1. The electric field inside the cluster and external negatively chirped laser pulse field with 40fs duration pulse, $6\times10^{14}$ $Wcm^{-2}$ laser intensity, and 825 nm wave length.

Figure 2. False color map of $T_e$ (a) and $Z_c$ (b)in Ar cluster at $t = 1.3\tau$ for an 825 nm wavelength pulse vs. pump intensity and pulse duration $\tau$=40fs.

Figure 3. (a). Time evolution of the electron density with $1\times10^{17}$ $Wcm^{-2}$ laser intensity, 40fs duration pulse and wave length of 825 nm. (b) The variations in the maximum electron density at the different laser pulse shapes in different laser intensities with 40fs duration pulse and wave length of 825 nm.

Figure 4. (a) Time evolution of the electron energy at the different laser pulse shapes with $1.5\times10^{17}$ $Wcm^{-2}$ laser intensity, 40fs duration pulse and wave length of 825 nm. (b) The variations in the maximum electron energy at the different laser pulse shapes in different laser intensities with 40fs duration pulse and wave length of 825 nm.

Figure 5. The variations in cluster size by changing the laser intensity at the different laser pulse shapes with 40fs duration pulse and wave length of 825 nm.

Figure 6. (a) Time evolution of the ion density of Ar cluster at different laser intensities of $3\times10^{14}Wcm^{-2}$, $6\times10^{14}Wcm^{-2}$ and $9\times10^{14}Wcm^{-2}$, (b) Time evolution of the ion density with the size of cluster.

Figure 7. The variations in maximum energy of electrons at the positively and negatively chirped laser pulse as a function of pulse durations with $2\times10^{17}Wcm^{-2}$ laser intensity and wave length of 825 nm.

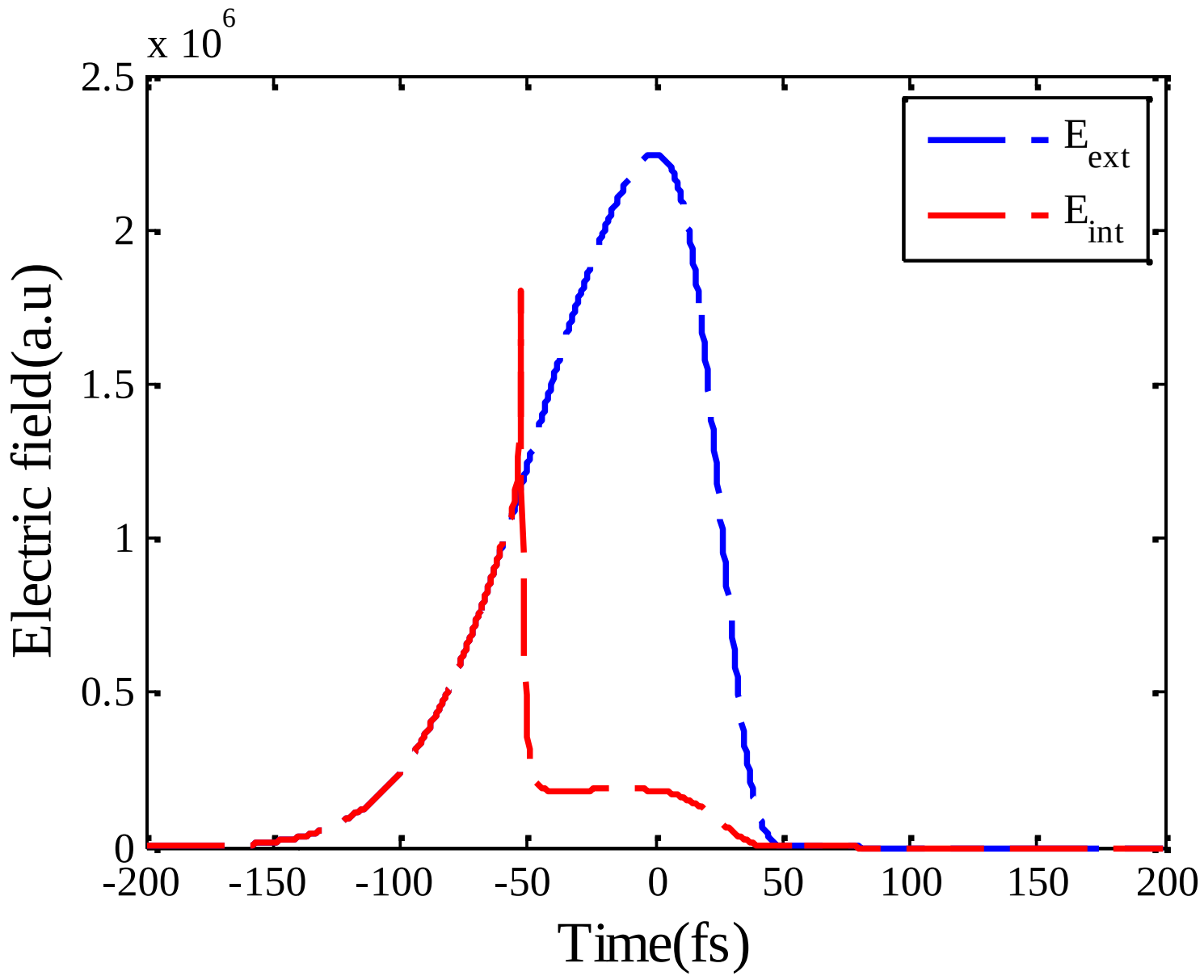

Figure 1

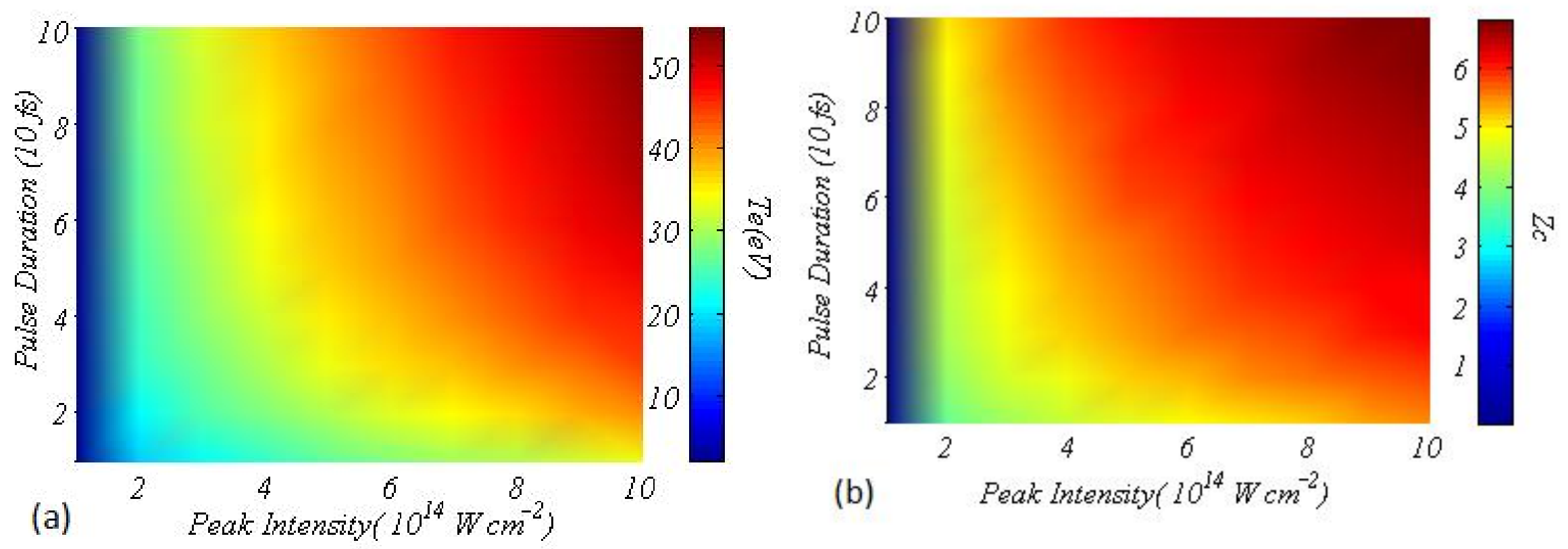

Figure 2

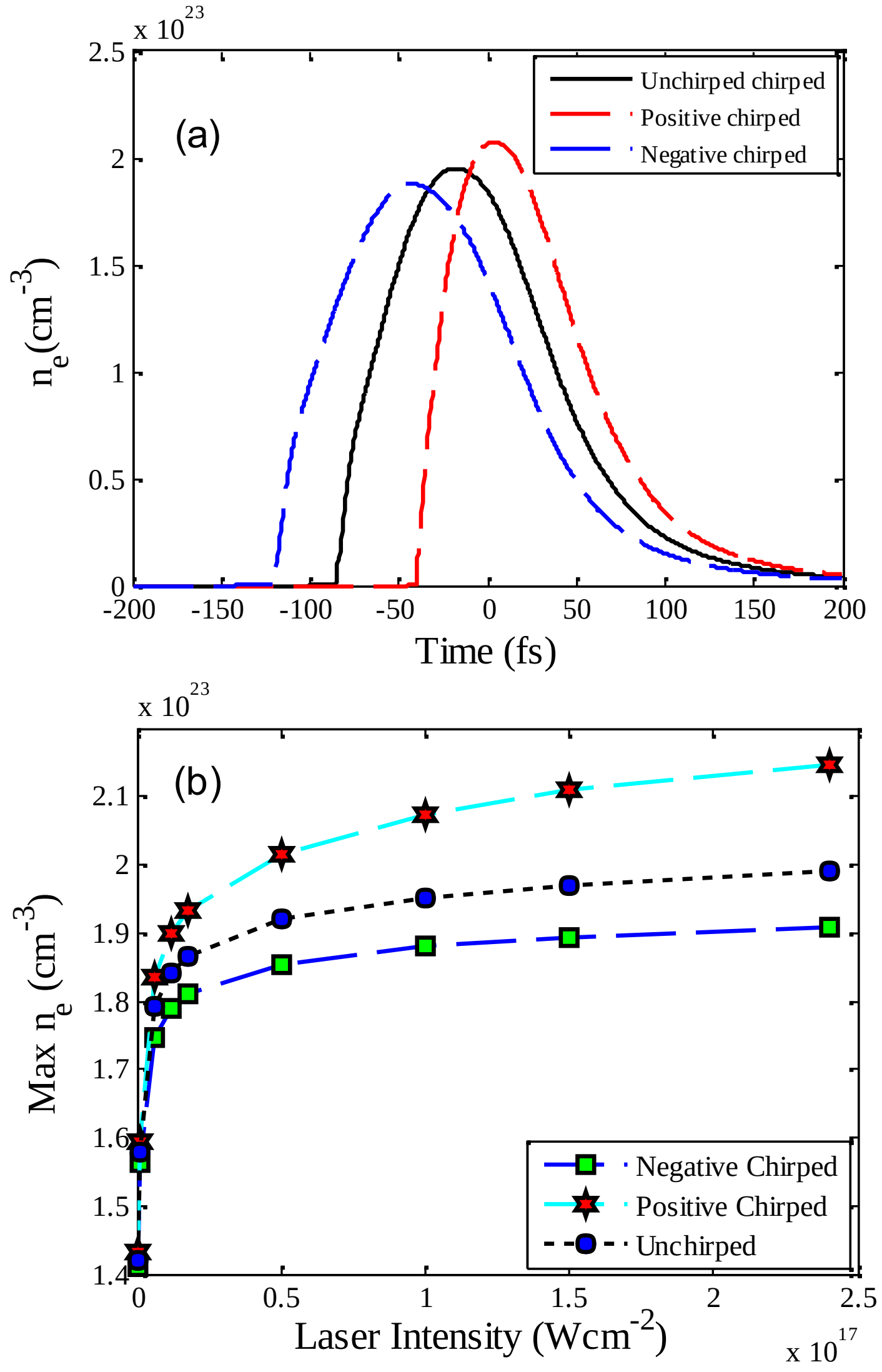

Figure 3

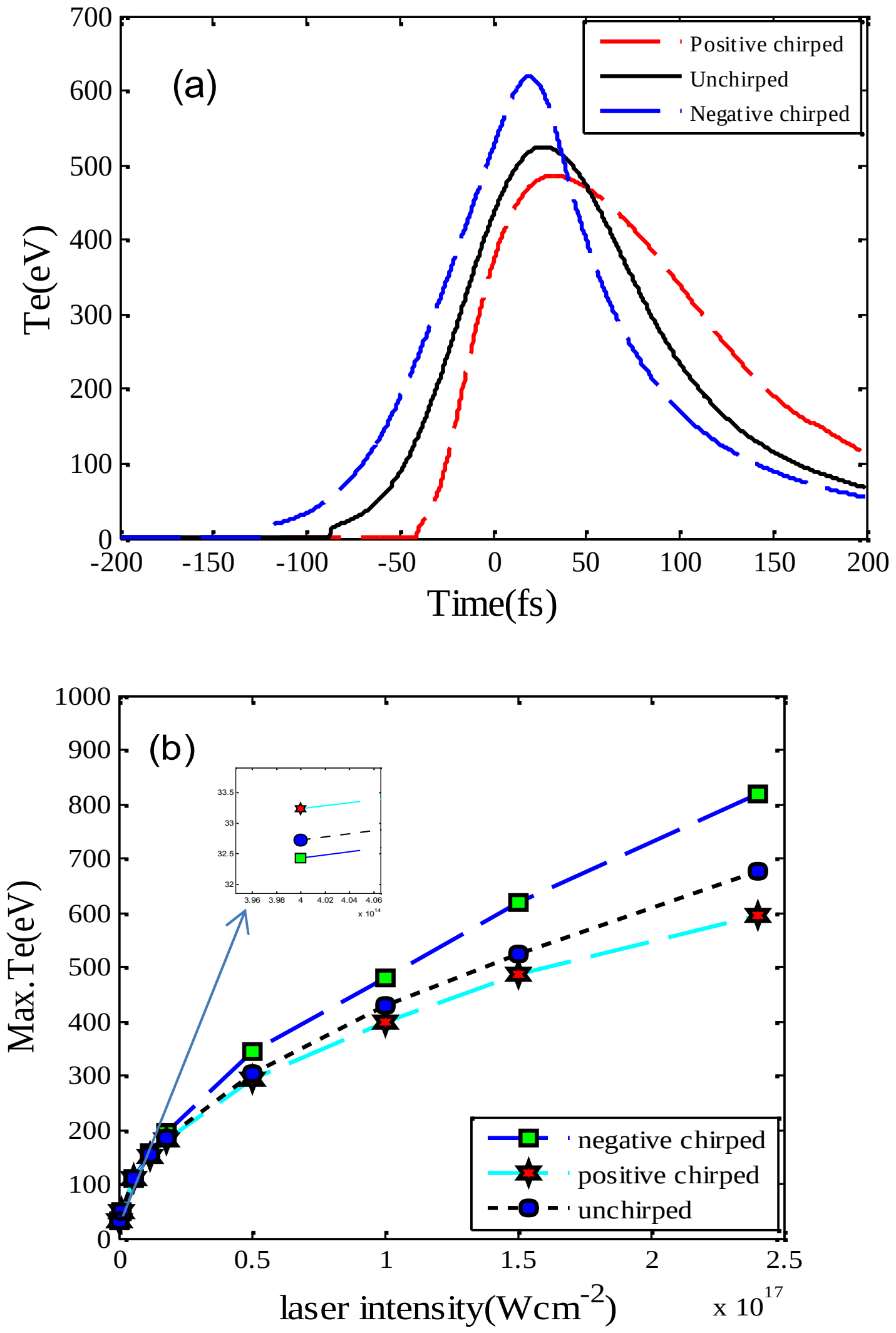

Figure 4

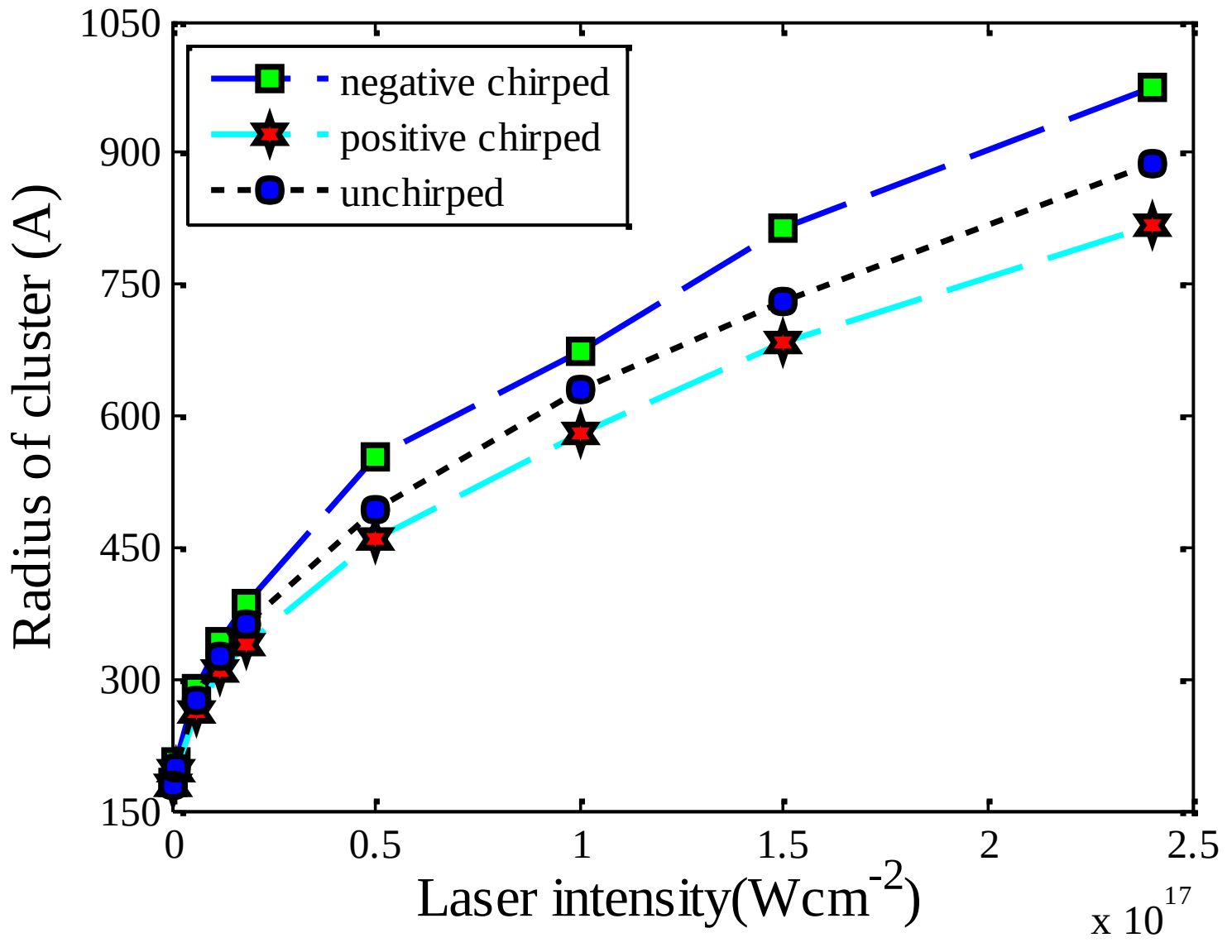

Figure 5

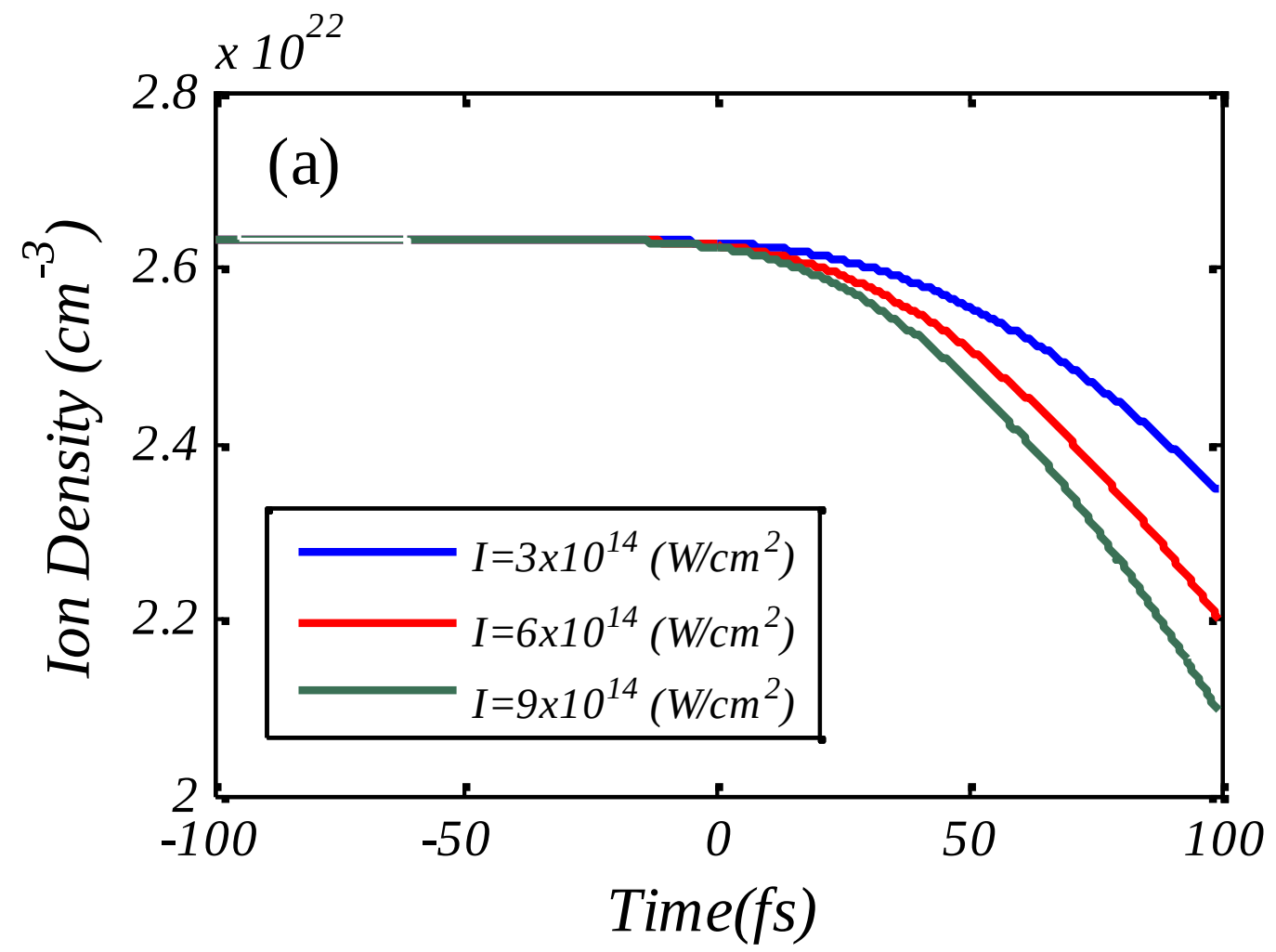

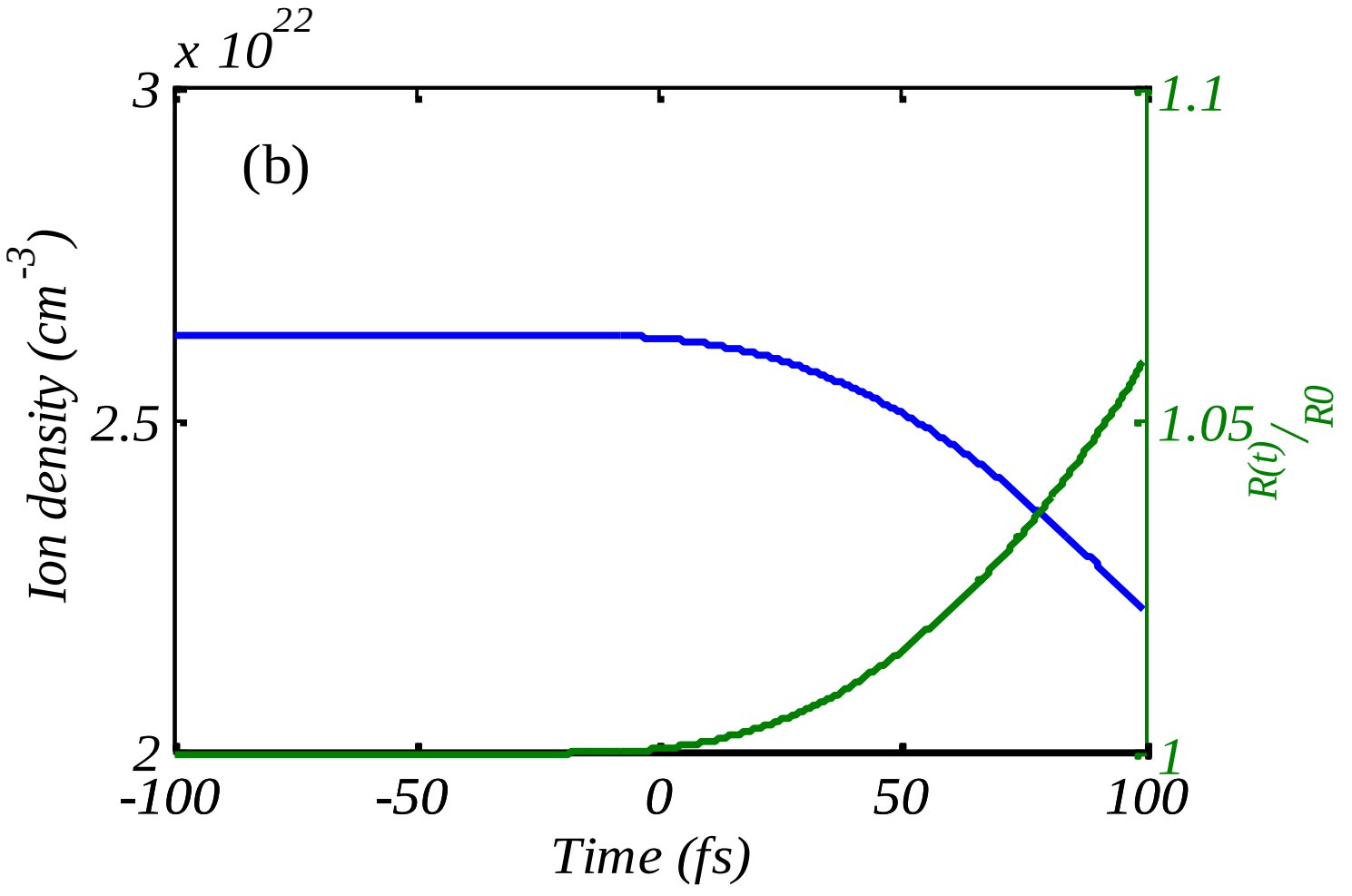

Figure 6

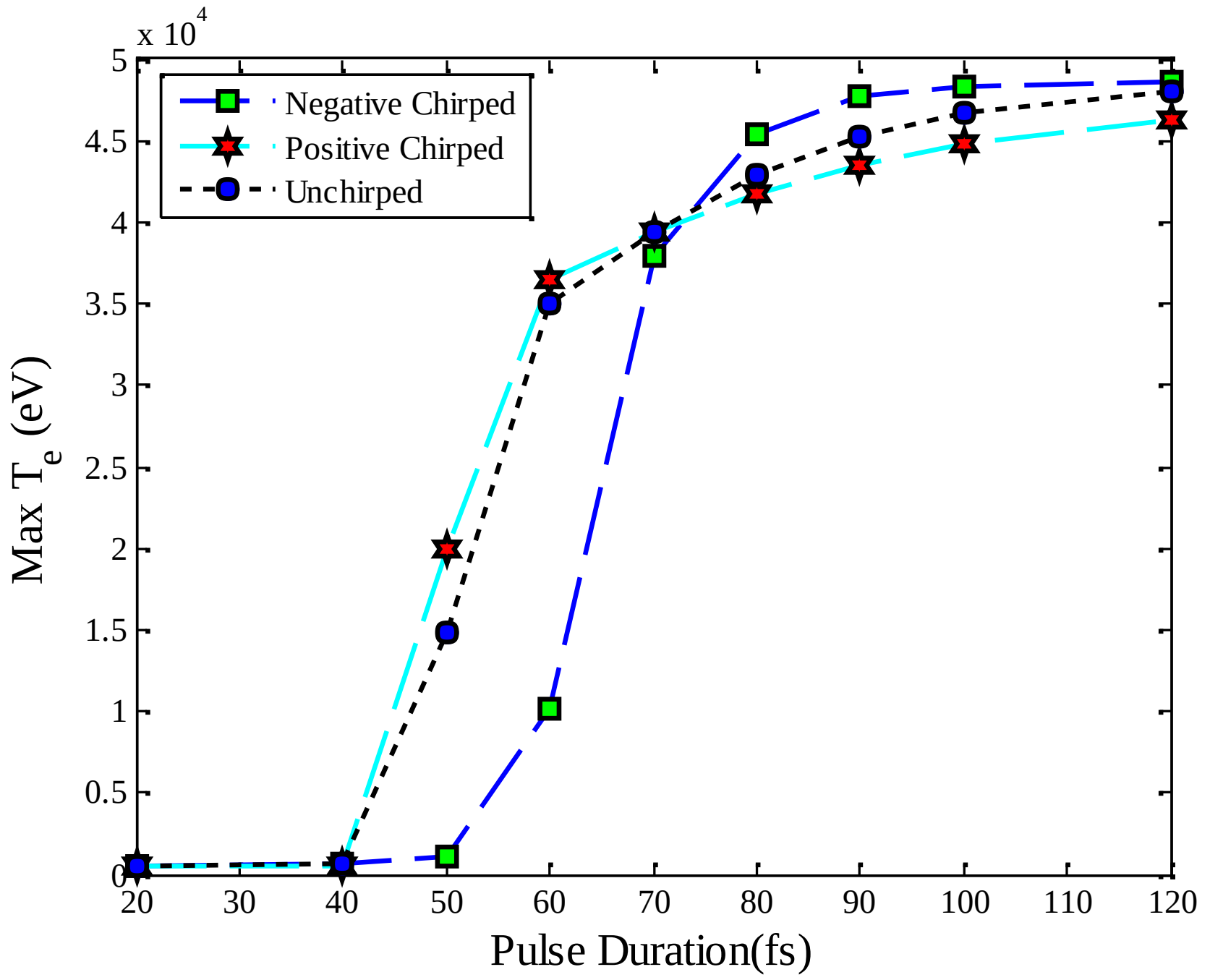

Figure 7